\documentstyle[prl,aps,epsf,twocolumn]{revtex}
\begin{document}   
\def\d{{\rm d}}

\title{Global and Local Aspects of Exceptional Points} 

\author{W.D.~Heiss }
\address{
Department of Physics, University of Stellenbosch, 7602 Matieland, South
Africa} 
\maketitle    
\begin{abstract} 
Exceptional points are singularities of eigenvalues and eigenvectors for
complex values of, say, an interaction parameter. They occur universally
and are square root branch point singularities
of the eigenvalues in the vicinity of level repulsions. 
The intricate connection between the distribution of exceptional points
and particular fluctuation properties of level spacing is discussed. 
The distribution of the exceptional
points of the problem $H_0+\lambda H_1$ is given for the situation of hard
chaos. Theoretical predictions of local properties of exceptional points 
have recently been confirmed experimentally. This relates to the specific
topological structure of an exceptional point as well as to the chiral 
properties of the wave functions associated with exceptional points. 
\end{abstract}
\medskip

\narrowtext

\section{Introduction}
We address the questions: which are the common properties of the 
{\sl quantum mechanical Hamilton operators} that give rise
to spectral properties that are ascribed to quantum chaos. If a matrix
representation of a Hamiltonian originating from a classically
chaotic analogous case is given, we explore the mathematical mechanism
that produces the special statistical features of the spectrum for the
particular parameter range where classical chaos is discerned.
While classical chaos has no intrinsic statistical
property, a further puzzling question is: why can a plain statistical
approach, with no physical input (such as GOE or GUE) \cite{boh}, reproduce 
the statistical properties of the spectrum so successfully?

We believe that the common root to the answer of these questions lies
in what is called the exceptional points\cite{kat} (EP) of an operator. Most
physical problems in quantum mechanics can be formulated by the
Hamiltonian $H_0+\lambda H_1$ where the parameter $\lambda $ can play the role
of a perturbation parameter, or it may serve to effect a phase
transition, or it may under variation steer the system from an ordered
into a chaotic regime. The EPs of the full operator
are the points $\lambda $  for which two eigenvalues coalesce. Here we exclude
genuine degeneracies of the self-adjoint problem, in other words, the
eigenvalues coincide for no real $\lambda $. The EPs occur
in the complex $\lambda $-plane. Note that the operator is not self-adjoint
for complex $\lambda $-values.

The definition of EPs is general and applies also
to operators in an infinite dimensional space, also when
the spectrum of the operator has a continuum part. In the present work
we restrict ourselves to finite $N$-dimensional matrices $H_0$ and $H_1$
as in this case the role of the EPs and the associated
Riemann sheet structure is thoroughly understood\cite{hesa,hest}. We do not
believe that restriction to matrices has a major impact on our
conclusions since virtually all the practical work even in connection
with quantum chaos is done in a finite dimensional matrix space.

The physical significance of the EPs is due to their
relation with avoided level crossing for real $\lambda $-values. The spectrum
$E_k(\lambda ),\, k=1,\ldots ,N$ has branch point singularities
at the EPs, in fact, any pair of the $N$ levels are 
generically connected by a square root branch point in the
complex $\lambda $-plane. If this happens near to the real
$\lambda $-axis, a level repulsion will occur for the two levels for real
$\lambda $-values. Globally, all the EPs determine
the shape of the whole spectrum. There is a nice analogy to the more
widely known connection between the singularities being poles of the
scattering function and the shape of the cross section: in a similar
way as the positions of the poles including their statistical properties
determine the measurable cross section, the EPs
determine the shape of the spectrum and in particular the occurrences
of avoided level crossings. The distribution of the
EPs will therefore determine the fluctuation properties of level
spacing.

The positions of the EPs are fixed in the complex
$\lambda $-plane and are determined solely by $H_0$ and $H_1$. For large
matrices it is prohibitive to determine the positions of the
EPs. However, it is possible to determine the
distribution reasonably well from the knowledge of the two operators.
Generically, a high density of EPs is a
sufficient prerequisite for the occurrence of quantum chaos if they
are randomly distributed according to a specific distribution function.

In the following section we recapitulate the basics about 
EPs. Section three
presents matrix models to exemplify the distribution of
EPs and level spacing fluctuations. Section four presents two recent
experiments where (i) the local topological structure of an
EP has been shown to
be a physical reality and (ii) where the chiral property of the wave
functions at the EP has been demonstrated. Section five presents a summary and
outlook.

\section{Exceptional points and unperturbed lines} 
Avoided level crossing is always associated with
EPs\cite{hesa,sh,hesa2} wherever they occur for the levels
$E_k(\lambda )$ of the Hamiltonian $H_0+\lambda H_1$.  We
give an elementary example for illustration and briefly list the
essential aspects with regard to EPs.

Consider a two dimensional matrix problem where $H_0$ is diagonal with
eigenvalues $\epsilon _1$ and $\epsilon _2$, while $H_1$ is represented in the
form \begin{equation} H_1=U\cdot D \cdot U^{-1}. \end{equation}
Here, the diagonal matrix $D$ contains the eigenvalues $\omega _1$ and
$\omega _2$ of the matrix $H_1$ and $U$ is the rotation
\begin{equation} U=\pmatrix{ \cos \phi & -\sin \phi \cr
    \sin \phi & \cos \phi }. \label{phas} \end{equation}
The eigenvalues of the problem $H_0+\lambda H_1$ are
\begin{equation} E_{1,2}(\lambda )={\epsilon _1+\epsilon _2+\lambda(\omega _1+\omega _2)\over 2}\pm R
 \end{equation} where
\begin{eqnarray} R&=&\biggl\{ \bigl({\epsilon _1-\epsilon _2 \over 2}\bigr)^2 
+\bigl({\lambda (\omega _1-\omega _2)\over 2}\bigr)^2 \cr
&+&{1\over 2}\lambda (\epsilon _1-\epsilon _2)(\omega _1-\omega _2)
\cos 2\phi \biggr\}^{1/2}. \end{eqnarray}
Clearly, when $\phi =0$ the spectrum is given by the two lines
\begin{equation} E_k^0(\lambda )=\epsilon _k+\lambda \omega _k \qquad k=1,2  \end{equation}
which intersect at the point of degeneracy $\lambda =-(\epsilon _1-\epsilon _2)/
(\omega _1-\omega _2)$. When the coupling between the two levels is turned
on by switching on $\phi $ the degeneracy is lifted and avoided level
crossing occurs. Now the two levels coalesce in the complex $\lambda $-plane
where $R$ vanishes which happens at the complex conjugate points
\begin{equation}\lambda _c=-{\epsilon _1-\epsilon _2 \over \omega _1-\omega _2}\exp (\pm2i\phi ).
\end{equation}
 At these points, the two levels $E_k(\lambda )$ are connected by a square
root branch point, in fact the two levels are the values of one analytic
function on two different Riemann sheets.

These considerations carry over to an $N$-dimensional problem\cite{hesa}.
The diagonal matrix $H_0$ contains the elements $\epsilon _k$ and
$D$ the elements $\omega _k,\,k=1,\ldots ,N$; the
matrix $U$ is now an $N$-dimensional rotation which can be
parametrized by $N(N-1)/2$ angles. (In the quoted paper a
parametrization was chosen so that $U$ is unity when all angles are
zero.) The EPs are
determined by the simultaneous solution of the equations
\begin{eqnarray}
\det(E-H_0-\lambda H_1)&=0  \cr
 {\d \over \d E}\det(E-H_0-\lambda H_1)&=0. \end{eqnarray}
There are generically $N(N-1)$ solutions which occur in complex
conjugate pairs in the $\lambda $-plane. At those points the $N$ levels
$E_k(\lambda )$ are connected in pairs by square root branch points when
they are analytically continued into the complex $\lambda $-plane. Since
the positions of the singularities determine the shape of the spectrum,
and in particular the fluctuation properties, a closer analysis is
indicated. As is exemplified in the next section, the crucial condition
for the occurrence of level statistics ascribed to quantum chaos is
a high density of EPs in the complex plane
within a small window of real $\lambda $-values.

To get an idea about the density of EPs we introduce
the concept of {\sl unperturbed lines}. Clearly,
when $U$ is the unit matrix (all angles are zero), the
spectrum of $H_0+\lambda H_1=H_0+\lambda D$ is given by the lines $\epsilon _k+\lambda \omega _k$
with $k=1,\ldots,N$. The $N(N-1)/2$ intersection points of the $N$ lines
depend on the relative order of the numbers $\epsilon _k$ and $\omega _k$. If
both sequences are in ascending order, all intersections occur at
negative $\lambda $-values; conversely, if one sequence is ascending and the
other descending all intersections occur at positive $\lambda $-values. In
general, the order which is appropriate for the actual problem, is
expected to lie between the two extremes. To find out the appropriate
order we are guided by the asymptotic behavior of the levels
$E_k(\lambda )$ of the full problem. For large values of $\lambda $ the leading
terms are given by
\begin{equation}E_k(\lambda )=\lambda \omega _k + \alpha _k + \ldots \label{lin}
\end{equation}
where the dots stand for first and  higher order terms in $1/\lambda $.
Neglecting these terms, Eq.\ref{lin} yields just the unperturbed
lines with the appropriate association of slopes $\omega _k$ and
intercepts $\alpha _k$. From perturbation theory we find the latter
to be the diagonal elements of the rotated $H_0$, {\it viz.}
\begin{equation} \alpha _k=(U^{-1}\cdot H_0\cdot U)_{k,k}\, .\end{equation}

\section{Distribution Function}
We begin with the distribution function of the intersection points of the
unperturbed lines as this can be derived rigorously. Assuming a uniform
random distribution for the $\omega _k$ and the $\epsilon _k$ (the eigenvalues 
of $H_1$ and $H_0$, respectively), the intersection points
$\lambda _{i,k}=-(\epsilon _i-\epsilon  _k)/(\omega _i-\omega _k)$ are
distributed according to  \cite{distr}
\begin{equation} P(x)={{\rm const}\over x^2}. \label{distr} \end{equation} 
The fact that $P(x)$ cannot be normalized
is due the fact that we assumed a uniform distribution of the $\omega _k$
over the whole range of the real numbers; for a finite sample there will
therefore be a deviation of $P(x)$ around $x=0$ from the form given in 
Eq.(\ref{distr}) as no $|\omega _k|$ is larger than some finite large number 
and hence an arbitrary small value of $1/\omega $ is unlikely to occur.

The intersection points are the points of degeneracies of $H_0+\lambda H_1$
as long as $U$ is equal to the identity matrix. We now gradually switch on
the interaction between the levels by switching on the angles randomly of
the random orthogonal matrix $U$ \cite{distr} but keeping initially 
the interval from which the angles are chosen very small. In this 
way the degeneracies become level repulsions
as the EPs start moving out into the complex 
$\lambda $-plane, a complex conjugate pair from each degeneracy. 
We conjecture that their distribution function remains
unchanged when they move out into the plane. For small values of the
mixing angles in $U$ they will of course be concentrated around the
real axis. The distribution function is now a function of two variables, 
say the real and imaginary part of a complex number. We
parametrize this number by $r\exp (i\alpha )$. For a fixed 
angle $\alpha $ the EPs  are distributed according to
\begin{equation} P(r)={{\rm const}\over r^2}. \label{distrr} \end{equation} 
This has been confirmed numerically in numerous cases. 
If the mixing angles of $U$ are very small, there is an obvious
dependence on $\alpha $ as the EPs cluster around the real 
axis; yet for fixed $\alpha $ the distribution law is always given by
Eq.({\ref{distrr}). Moreover, for the nearest neighbor distribution (NND) of the
energy levels it turns out that a proper Wigner distribution is obtained
only when the EPs have fanned out into the plane in such a way
that their distribution becomes independent of the angle $\alpha $. 
In Fig.1 it is illustrated how the EPs fan out 
into the plane when the mixing angles of $U$ are turned on.

\begin{figure}
\epsfxsize=2.2in
\centerline{
\epsffile{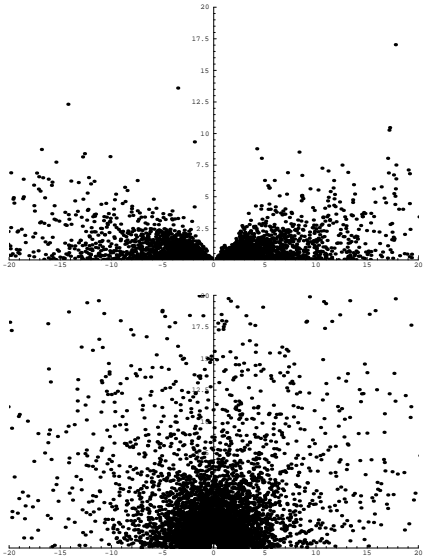}}
\vglue 0.15cm
\caption{ Exceptional points as they move into the complex $\lambda $-plane 
when the angles in $U$ are switched on. The lower part illustrates hard chaos.
}
\label{fig1}
\end{figure}

The law of Eq.(\ref{distrr}) is verified numerically even if the EPs
originate out of a degeneracy of high multiplicity. In Fig.2 the emergence
of EPs out of a point of degeneracy is illustrated. Here,
Eq.(\ref{distrr}) is obtained once they fill the plane in an isotropic way,
that is when the distribution has become independent of $\alpha $.

To summarize: hard chaos, that is a Wigner distribution for the NND of the 
energy levels, is associated with a distribution of EPs in
the complex plane that is independent of the angle $\alpha $ of the
complex number $r \exp (i\alpha )$ and depends only on the distance $r$
according to Eq.(\ref{distrr}). If an $\alpha $-dependence prevails, the
NND is not Wigneresque. In the limiting case ($U\equiv $ identity) of the example 
used above the NND is a Poisson distribution.

\begin{figure}
\epsfxsize=2.2in
\centerline{
\epsffile{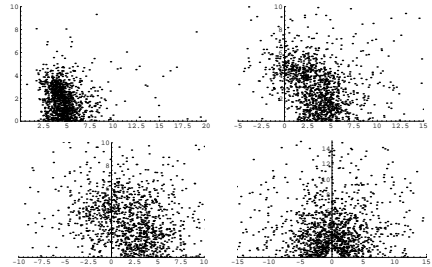}}
\vglue 0.15cm
\caption{ Exceptional points as they move into the complex $\lambda $-plane 
when the angles in $U$ are switched on. Here a point of degeneracy at $\lambda =5$ 
is the starting point. In the figure on the right bottom this information is
completely lost and hard chaos prevails.
}
\label{fig2}
\end{figure}

We only mention that in special non-generic cases the EPs
are arranged in a geometrically ordered pattern like in the
integrable Lipkin model \cite{lip}. However, a small perturbation
leads to the generic situation as discussed above \cite{hesa2}.

\section{Local Properties of Exceptional Points}
The topological structure of the square root branch point associated
with an EP has been shown to be a physical reality in a 
recent experiment \cite{dembo}. As a particular consequence it has been
established experimentally that the phases of the wave functions that 
take part in the coalescence of the two energy levels show a phase behavior
that is {\sl distinctly different} from that of a usual degeneracy, i.e.~from that
at a diabolic point \cite{berry}. We emphasise that one major signature, contrasting
an EP from a degeneracy, is the lack of two independent eigenfunctions; there is
in fact only one eigenfunction: the EPs are the points where the Jordan form
of the operator does not give a diagonal matrix \cite{kat}. 
This is sometimes overlooked in the literature \cite{Latinne,Rotter}. 
They can also occur in the continuum as the coalescence of two
resonances \cite{Mondragon}.
The crucial experiment \cite{dembo} yielded three major results which have been
predicted in \cite{heiss}:
\begin{enumerate} 
\item  If a loop is performed in the $\lambda$-plane around the EP,  
the eigenenergies $E_1$ and $E_2$ are interchanged.  
\item The wave functions $|\psi _1\rangle$ and  
$|\psi_2\rangle$ are interchanged by the loop and, in addition, one  
of them changes sign. In other words, a loop in the $\lambda $-plane 
transforms the pair $\{\psi_1,\psi_2\}$ into $ \{-\psi_2,\psi_1\}$.  
Therefore the two possible directions of looping  yield  
different phase behavior. In fact, encircling the 
EP a second time in the same direction, we obtain 
$\{-\psi_1,-\psi_2\}$ while the next loop yields 
$\{\psi_2,-\psi_1\}$ and only the fourth loop restores 
the original pair. It follows that by going in the opposite  
direction, one finds after the first loop what is obtained after three 
loops in the former direction. This finding confirms the fourth root
character of singularity for the wave functions.
\item The eigenvalues $E_1, E_2$ have been studied as 
functions of $\lambda$ for two paths that were not closed. One path 
was just above, the other one just below $\lambda _c$. The results were 
different. Calling the real part of an eigenvalue the resonance energy
and its imaginary part the  resonance width. On one of the 
paths, the widths cross while the resonance energies avoid each other. 
On the other path, the resonance energies cross 
while the widths avoid each other. 
\end{enumerate} 
 
We conclude the discussion with a further important local
property of EPs. The single and unique (up to a global factor)
wave function at the EP
has always a specific chiral behavior \cite{heha}. A recent experiment
at the TU-Darmstadt \cite{demboch} has directly confirmed this feature.
In similar context this has been discussed for acoustic waves in a 
medium \cite{shuv} and indirectly observed in optics \cite{panch}.
The latter has been explained in terms of EPs in
\cite{ber2}. The essential finding of \cite{heha} is the
unique form of the
wave function at the EP being {\sl always} of the form
\begin{equation} |\psi _{{\rm EP}}\rangle = 
|\psi _1\rangle \pm i |\psi _2\rangle  \label{supp}
\end{equation}
where the plus or minus sign refers to a specific EP.
The $|\psi _i\rangle $ are the two wave functions that coalesce
at the EP. No other superposition is possible, irrespective
of a particular physical situation such as driving the
dissipating system. For a time dependent problem
this signals chiral behavior. If the two wave functions relate to 
different parities or to different linear polarizations, the
superposition is obviously chiral; in the latter case it is a
circularly polarized wave of specific orientation. We recall
that here $H_0$ and $H_1$ are assumed to be hermitian. If
the two operators are non-hermitian there is still a unique
superposition but Eq.(\ref{supp}) has to be modified \cite{herec}.

\section{Summary and Outlook}
Exceptional points are a fascinating subject of theoretical physics.
As they are the only singularities of the spectrum for a matrix
problem, they `make' the spectrum, so to speak. They are directly
associated with level repulsion. As a consequence, their statistical
properties relate directly to that of the spectrum itself. Integrable
systems give rise to a geometrical ordering while chaotic systems
relate to disordered arrangements, yet with a specific distribution 
function.

The local behavior is fascinating on its own.
The topological structure of a square root branch point is a physical
reality with all its consequences for the wave functions. In
addition, with its 
intrinsic chiral behavior it may even hold some promise to shed light
on the ubiquitous left-right asymmetry of our macroscopic world.
On a speculative note: the kinematic relation $E=\pm \sqrt{\vec p ^2+m^2}$ 
bears all properties of an EP at $|\vec p|=\pm im$. 
It was this relationship that led Dirac to his famous equation 
including spin and the properties of chirality.

\end{document}